\documentclass[aps,superscriptaddress,twocolumn,showpacs,preprintnumbers,amsmath,amssymb,showkeys,textcomp]{revtex4}
\usepackage[cp1251]{inputenc}
\usepackage{amssymb,amsmath,wasysym}
\usepackage[mathscr]{eucal}
\usepackage{graphicx}
\usepackage{longtable}
\usepackage[dvips]{hyperref} 
\usepackage[cp1251]{inputenc}
\usepackage[english]{babel}
\usepackage{amssymb,amsmath}
\usepackage{amstext} 
\usepackage[dvips]{hyperref}
\usepackage[mathscr]{eucal}
\usepackage{longtable}
\usepackage{graphicx}
\begin{document}
\title{To Photon Concept and to Physics of Quantum Absorption Process}
\author{Dmitri Yerchuck (a), Yauhen Yerchak (b),   Alla Dovlatova (c), Vyacheslav Stelmakh (b), Felix Borovik   (a)\\
\textit{(a) - Heat-Mass Transfer Institute of National Academy of Sciences of RB, Brovka Str.15, Minsk, 220072,  dpy@tut.by \\(b) - Belarusian State University, Nezavisimosti Ave., 4, Minsk, 220030, RB \\(c) M.V.Lomonosov Moscow State University, Moscow, 119899}}
\date{\today}
\begin{abstract} The status of the photon in the modern physics was analysed. Within the frames of the Standard Model of particle physics the photon is considered to be the genuine elementary particle, being to be the messenger of the electromagnetic
interaction to which are subject charged particles. In contrast, the experts in quantum electodynamics (in particular, in quantum optics) insist, that 
the description of an photon to be the particle is impossible. The given viewpoint was carefully analysed and its falseness was proved. The expression for a photon wave function is presented. So, the status of the photon in
quantum electodynamics was restored. The physics of a  quantum absorption process is analysed. It is argued in accordance with Dirac guess, that the photon revival takes place by its absorption. Being to be a soliton, it seems to be keeping safe after an energy absorption in a pinned state, possessing the only by spin. It is shown, that  the time  of the transfer of absorbing systems in an excited state is finite and moreover, that it can govern the stationary signal  registered. The given result is significant for the all stationary spectroscopy, in which at present the transfer of absorbing systems in an excited state is considered to be instantaneous. 
\end{abstract}
\maketitle 
\section{Introduction and Background}

The studies of optical properties of a number of condensed matter systems have shown, that there are phenomena, which can be explained the only within the frames of the concept of quantized electromagnetic field (EM-field) interacting with given systems, which also have to be described quantum-mechanically.
 
The quantization theory of EM-field  was proposed for the first time still at the earliest stage of quantum physics in the works \cite{Born},\cite{Born_Heisenberg}. In given works quantum theory of dipole radiation was considered and the energy fluctuations in radiation field of blackbody have been calculated. The idea of Born-Jordan EM-field quantization is regarding of EM-field components to be matrices. Given idea, being to be mathematically correct does not give any indications on the nature and on the character of the structure of EM-field, althogh it points out on its discreteness. At the same time quite another idea - to set up in the 
correspondence to each mode of radiation field the quantized harmonic oscillator, was proposed for the first time by Dirac  \cite{P.Dirac} and it is widely used in quantum electrodynamics (QED) including quantum optics,  it is canonical quantization. Given idea has fundamental physical base. It allowed to unite the corpuscular and wave properties of the light, that is, to explain the quantum phenomenon of  corpuscular and wave dualism.

Let us give a brief review of the  history of the experimental confirmation of the necessity to consider the EM-field to be quantised. A.Einstein was the first, who
proposed to describe the EM-field to be quantised on the base of an analysis of the photoeffect phenomenon \cite{Einstein}, \cite{Einstein_A}. Let us remember, that the given phenomenon was discovered by Hertz in 1887, was studied by Stoletov (1888), Lenard and Tomson (1889), however, the only A.Einstein  have explained the main regularities of the photoeffect in the suggestion that that the energy of EM-field has to be quantised in 1905, that is, long before the appearance of first works on the theory of  the EM-field quantization  \cite{Born},\cite{Born_Heisenberg},\cite{P.Dirac}. Let us 
regard the reasons of A.Einstein proposal in more details. There were explained the following main properties: 1) the independence of the maximal  kinetic energy 
of photoelectrons on the light intensity, 2) the linear dependence of the maximal  kinetic energy 
of photoelectrons on the light frequency and 3) the existence of the minimal light frequency (threshold frequency). Let us remark, that  the linear dependence of the maximal  kinetic energy 
of photoelectrons on the light frequency can be explained on the base of ealier Planck idea \cite{Planck}  of the quantisation of the energy of light absorbing atoms, that is, that the absorbing atoms scoop the energy by the portions from the continuos EM-field medium, in other words, by Planck consideration of  the processes, in which the energy of the  light was the continuos quantity. At the same time, the independence of the maximal  kinetic energy of photoelectrons on the light intensity  and the existence of the threshold light frequency could be explained the only on the base of the quantisation of the energy of the light, just that has been done by A.Einstein \cite{Einstein}. 

Let us concern the basic quantisation paradigm itself introduced in the physics by Planck. We have to remark for the sake of a historical truth that the idea of  absorption or emmission of the energy of EM-field with material oscillators by discrete portions was in fact proved by Planck, that is, not postulated (the words of the type "Max Planck postulated the discretness, or "Planck postulated, that the oscillators in the walls of the cavity can only absorb or emit radiation in discrete units" and so on are often occured in the scientific literature and textbooks, see, for instance, \cite{Schleich}). Really, the proof was based the only on the following: "Die Hypothese, welche wir jetzt der weiteren Rechnung
zu Grunde legen wollen, lautet folgendermassen: Die
Warscheinlichkeit $W$ daf\"{u}r, dass die $N$ Resonatoren insgesamt
die Schwingungsenergie $U_N$, besitzen, ist proportional der Anzahl $\mathfrak{R}$ aller bei der Verteilung der Energie $U_N$ auf die N Resonatoren
m\"{o}glichen Complexionen; oder mit anderen Worten:
irgend eine bestimmte Complexion ist ebenso wahrscheinlich,
wie irgend eine andere bestimmte Complexion.["The hypothesis, which we now will base  on the subsequent calculations is formulated in the following way: the probability $W$ of that, that $N$ oscillators taken together possess by the vibration energy $U_N$ is proportional to the number $\mathfrak{R}$ of all possible Complexions  by the distribution of the  energy $U_N$ on $N$ oscillators or in other words - any definite Complexion haa the same probability with every other Complexion (the term "Complexion" Max Planck elucidates: "Die Verteilung der $P$ Energieelemente
auf die $N$ Resonatoren nur auf eine endliche ganz
bestimmte Anzahl von Arten erfolgen kann. Jede solche Art
der Verteilung nennen wir nach einem von L. Boltzmann f\"{u}r
einen \"{a}hnlichen Begriff gebrauchten Ausdruck eine "Complexion". [The distribution of the $P$ energy elements {that is energy quanta in modern terminology}  on $N$ oscillators can be produced by the only finite numbers of the ways. Each given way we will call in accordance with the similar notion, used by L. Boltzmann a "Complexion"). Planck writes further : "Ob diese Hypothese
in der Natur wirklich zutrifft, kann in letzter Linie nur
durch die Erfahrung gepr\"{u}ft werden [Whether the given hypothesis is in the Nature really taking place, it can be verified first of all by an experiment]".
Concerning the proof itself. Planck \cite{Planck}  has derived two independent expressions for the entropy of the absorption or emission  processes by material oscillators. They are the following
\begin{equation}
\label{eq1ab}
\begin{split}
S_N = k N[(1 + \frac{U}{\epsilon})\log(1 + \frac{U}{\epsilon}) -  \frac{U}{\epsilon}\log(\frac{U}{\epsilon})],
\end{split}
\end{equation}
and 
\begin{equation}
\label{eq2ab}
\begin{split}
S = f(\frac{U}{\nu}),
\end{split}
\end{equation}
where  $\epsilon$ is energy element (energy quantum) value, $S_N$ is the entropy of $N$ independent oscillators, S is the entropy of an individual oscillator, $\nu$ is oscillator frequency. It is seen from (\ref{eq2ab}) that the entropy of oscillators in any diathermic medium is the function of the only one argument $\frac{U}{\nu}$, that was accentuated by Planck. It is followed then by the comparison of the expressions (\ref{eq1ab}) and (\ref{eq2ab}) mathematically strictly the relation

\begin{equation}
\label{eq3ab}
\begin{split}
\epsilon = h\nu,
\end{split}
\end{equation}
where $h$ has to be the universal constant.  Its value, equaled to $6.55\times{10^{-27}}$ $erg\times{s}$  was also determined by Planck \cite{Planck}.
It is the main result of the paper \cite{Planck}, owing to which the new, quantum era in the physics was started.

The above cited hypothesis, used by  Planck,  can also be grounded. Really, the equal probability of Complexion distribution means that the absorption (emission) process for every from $N$ independent oscillators, distributed in a space, will be described by the same characteristics independently of the individual oscillator space location and the starting absorption (emission) time moment. So, we obtain that the equal probability of Complexion distribution will be realised, when the free  space and time are homogeneous. In other words, the proof of the   Planck hypothesis aforecited results from the homogeneity of Minkowsky space. The homogeneity of the space and times was grounded slightly later after publishing in 1918 by N\"{o}ter \cite{Noether} her famous theorem, which being to be applied to the symmetry study of the time and space has indicated on their homogeneity, since the only in the given case the experimentally very good confirmed conservation laws of the energy and impulse in the mechanics were valid. It is understandable why the given small step was remained in 1900 to the entirely mathematically strict proof of the relation (\ref{eq3ab}).

We wish to accentuate that the Einstein concept of light quanta to be the particles was the new paradigm, introduced in the physics for the first time basing on experimental results, that is, experimentally grounded. In fact, it was the recovery of I.Newton concept on the corpuscular nature of  the light on the new scientific level. Let us remark that the representation on the corpuscular nature of the light  was considered  by I.Newton the only on the intuitive level without any experimental grounds and we have to give his ingenious insight due.

The second physical phenomenon, indicating on the quantum nature of EM-field is Compton effect, which was experimentally thoroughly studied for the first time by Compton \cite{Compton} in 1922. Compton \cite{Compton_A} and independently Debye have  proposed the elementary theoretical explanation of the phenomenon observed on the base of corpuscular nature of X-rays. The analysis has shown that the impulse of EM-field is also quantised. In other words, light quanta, possessing by the energy $\hbar\omega$ possess also by the impulse, corresponding to the given energy, equaled to $\vec p  = (\hbar\omega/c)\vec e$, where $\vec e$ is the unit vector in the propagation direction. 
 
It is interesting that the independent conclusion on the the impulse quantisation of light quanta, the value of which is dependent on the energy quantum value is resulted from experiments of P N Lebedev \cite{Lebedev} on the pressure of light, which were performed already in 1901, that is substantially earlier in comparison with Compton experiments. Really, Lebedev's experiments have verified with high certainty the existence of the pressure of light, described by the following expression
\begin{equation}
\label{eq4ab}
\begin{split}
P = \frac{E}{c},
\end{split}
\end{equation}
which means that to any portion $E$ of the light energy is set up in the correspondence the mechanical impulse $P$. Taking into account the Einstein relation for the energy of a light quanta, we obtain
\begin{equation}
\label{eq5ab}
\begin{split}
P = \frac{h\nu}{c},
\end{split}
\end{equation}
in the full correspondence with independent Compton results.

Let us remember, that light quanta  were called photons
by Lewis in the spirit of  I.Newton  corpuscular concept  slightly later, in 1926 \cite{Lewis}.

There were revealed very interesting properties of the light flux in the experiments on the interference and diffracion of light described in \cite{Dempster}. it has been found that at very low intensities an interference pattern is not appeared. At the same time the atoms of free silver are appeared on a photographic  plate in the place of photon falling. They represent themselves an embryo, which is much smaller than a light wave length. The particle properties of the light show up in the birth of free silver atoms on a photographic  plate, one by one.  At the same time, when the light intensity is rather large an   interference pattern has been shown up. The attempt to explain the given results was undertaken in the book \cite{Braginsky}. The authors write: " In the interference process (e.g. in two-slit experiment) the photon must have been influenced by the locations of both slits, since the interference pattern depends on the distance between them". From hence the authors conclude that "the photon must have occupied a volume larger than the  slit separation". The authors continue further: "On the other hand, when it fell on the photographic  plate the photon must have become localized in the tiny volume of the silver embryo" The explanation of the given fact according to \cite{Braginsky} consist in the presence of "collapse of wave function" and "reduction of the wave packet" processes. They conclude:"The wave properties of the photon show up in the fact that the probability of the collapse at a certain place on the photographic  plate (and the accompanying birth of a silver atom there) is proportional to the light intensity". 

The explanation above cited seems to be very vague. It is not understandable, how any elementary particle can have simultaneously very different sizes being to be not subjected to external effects. We take in mind that the presence of  the dependence of the probability of the photon size collapse on the light intensity, which was concluded in \cite{Braginsky}, is in fact the direct indication on the interaction between photons within the frames of the model proposed in \cite{Braginsky}. It  is well known, however, that all  quantum optics effects can be explained in the suggestion of noninteracting between themselves photons. However, even  on the assumption of some interaction between photons the very strong change in the photon size seems to be unreal. 

The results described in \cite{Dempster} is the good example for the display of the reality of rather complicated structure of EM-field, the Fermi liquid model of which is recently proposed in \cite{DAA} (see the next Section, where the   correct explanation of the aforedescribed results is represented on the base of the given model).

It is known at present a number of quantum optics phenomena, which can be described the only with  the regard for the  EM-field quantization, the main of which are briefly reviewed below.
The very interesting consequence of the EM-field quantization is the appearance of the vibrations which correspond to zeroth energy, so-called vacuum fluctuations. Vacuum fluctuations do not have  any classical analogue and they determine very many interesting quantum optics phenomena. For example, spontaneous emission  can be explained to be a result an atom "stimulation" by vacuum fluctuations. 

Let us remark that the results of fully quantum consideration are quite different from the results, obtained by means of the semiclassical theory of the interaction of EM-field with an atom, in which the atom is considered quantum-mechanicaly, but EM-field is considered clasically, even in the case when vacuum fluctuations are included fenomenologically into semiclassical consideration, see, for instance the the example of the task on quantum beats \cite{Scully}. 

The quantization of EM-field is necessary for the explanation along with a spontaneous emission phenomenon of other quantum optics phenomena - Lamb shift, Casimir effect, an entaglement of the states \cite{Scully}, \cite{Schleich}. It allows also to explain the value of linewidths of the lasers' emission,  to explain the observation of nonclassical squeezed states of EM-field and to give the correct desription of a complete statistics of laser photons \cite{Scully}, \cite{Schleich}.

 Further,
the QED model for a multichain coupled
 qubit system proposed in \cite{Dovlatova_Yearchuck} predicts that by a strong electron-photon interaction the quantum nature of  EM-field can
 become apparent in any stationary optical experiment. The conclusion is based on  a new quantum physics phenomenon — the space propagation of quantum Rabi oscillations, which  has been theoretically predicted  in \cite{Slepyan_Yerchak} for the systems with a strong electron-photon interaction. The notion of  the new quantum objects - rabitons, that is, the quasiparticles which have dual corpuscular-wave nature like to the well known other quantum objects, has been introduced in \cite{Slepyan_Yerchak}. It was theoretically predicted  in \cite{Dovlatova_Yearchuck} that the rabiton formation can give an essential contribution in the stationary spectral distribution of Raman scattering (RS) intensity, spectral distributions of infrared (IR), visible, or ultraviolet
 absorption, reflection, and transmission intensities. The given prediction was confirmed in \cite{Yerchuck_D_Dovlatova_A}, where the additional lines
 corresponding to the Fourier transform of the revival part of  the time dependence of the integral inversion were identified. In other words the lines, the appearance   
 of which in stationary RS  and IR measurements is determined by the
 formation and the propagation of quantum Rabi corpuscular-wave packets have been registered and/or identified in a number of quasionedimensional and two-dimensional carbon systems. Especially interesting that additionaql lines are observed the only in experiments, when the coherent state of the electronic system, interacting with an external EM-field is retained, and they disappear (in distincton from main usual lines) by the violation of the coherence, which was experimentally achieved by the adjustment of spectra registration conditions, corresponding to the stochastic regime \cite{Yerchuck_D_Dovlatova_A}. It seems to be appropriate to draw the attention, that the transition to the stochastic behaviour is characteristic for quantum systems. For instance, the authors of \cite{Shigehara} have theoretically considered the chaos induced by a quantization. They showed that two-dimensional billiards with point interactions inside exhibit a
chaotic nature in thecase of quantum systems, although their classical counterparts are non-chaotic. They indicate also, that quantum billiard considered is a natural starting point for examining
the particle motion in microscopic bounded regions and that
the rapid progress in the microscopic and mesoscopic
technology makes it possible to realize given  settings. So we can conclude, that the observation of the stochastic behaviour itself of the studied optical systems is the strong indication on the  necessity of a full quantum description of the processes in a joint system $\{$EM-field+matter$\}$.  

In spite of rapid development of quantum optics the base method for the theoretical descripion - canonical Dirac quantization method \cite{P.Dirac}, proposed in 1927, has remained  a very long time without any changes. At the same time, Dirac himself has accentuated, that the theory proposed does not sufficiently strict quantum-relativistic theory and the main weakness  of the theory is that the time is considered, being to c-number, instead of, rather than to consider it symmetrically with  the space coordinates \cite{P.Dirac}. Nevertheless, Dirac has concluded, that the theory proposed describes fairly satisfactorily the emission of a radiation and the reaction of the  radiation field  on the emitting system.
Dirac has used the same method  for the description of the quantum dispersion \cite{Dirac}. In succeeding years, many theoretical results have been obtained in quantum electrodynamics by using of Dirac quantization method, which were rather well agreeing with experimental data. The method was called canonical, and the impression arises, that the researchers in the field forgot without a trace the Dirac's remark, that the method proposed is in fact semi-quantum-relativistic, that is, being to be local relatively the time coordinate, it has the  global character relatively the space coordinates. Really, it is easily to find  an explicit form for the dependencies of operator functions   $\hat{a}_{\alpha}(t)$ and $ \hat{a}^{+}_{\alpha}(t)$ on the time by canonical quantization. It has been done in \cite{Dovlatova_Yerchuck}. They are the following
\begin{equation}
\label{eq30ab}
\begin{split}
&\hat{a}^{+}_{\alpha}(t) = \hat{a}^{+}_{\alpha}(t = 0) e^{i\omega_{\alpha}t},\\
&\hat{a}_{\alpha}(t) = \hat{a}_{\alpha}(t = 0) e^{-i\omega_{\alpha}t},
\end{split}
\end{equation}
where $\hat{a}^{+}_{\alpha}(t = 0), \hat{a}_{\alpha}(t = 0)$ are constant, complex-valued in general case, operators. 

Physical sense of operator time dependent functions $\hat{a}^{+}_{\alpha}(t)$ and $\hat{a}_{\alpha}(t)$ is well known. They are creation  and annihilation operator of the $\alpha$-mode photon in multimode EM-field.
They are continuously differentiable operator functions of time. It means, that the time of  photon creation (annihilation) can be determined strictly, at the same time operator  functions $\hat{a}^{+}_{\alpha}(t)$ and $\hat{a}_{\alpha}(t)$  do not curry any information on the place, that is, on space coordinates of a given event, confirming the space-global character of the given semi-quantum-relativistic quantization method.

Let us also remark, that the very similar viewpoint concerning the status of time-coordinate in quatum mechanics (QM) was advanced by Schr\"{o}dinger \cite{Schrodinger}:
 "Ich m\"{o}chte wiederholen, dass wir eine QM,
deren Aussagen nicht f\"{u}ur scharf bestimmte Zeitpunkte
gelten soIlen, nicht besitzen. Mir scheint,
dass dieser Mangel sich gerade in jenen Antinomien
kundgibt. Womit ich nicht sagen will, dass es der
einzige Mangel ist, der sich in ihnen kundgibt. Dass die "scharfe Zeit" eine Inkonsequenz innerhalb der QM ist und dass ausserdem, sozusagen
unabh\"angig davon, die Sonderstellung der
Zeit ein schweres Hindernis bildet f\"{u}r die Anpassung
der QM an das Relativit\"{a}tsprinzip, darauf
habe ich in den letzten Jahren immer wieder hingewiesen,
1eider ohne den Schatten eines branchbaren
Gegenvorschlags machen zu k\"{o}nnen" [I would wish to repeat, that we don't have QM, the  substance of which would be regarded to not strictly determined time moments. It seems to me, that the given disadvantage (demerit) is displayed in its contradictions. I don't wish to say, that the given demerit is only one, which is revealed in them. I have time and again pointed out in the last years, \cite{SchrodingerA} \cite{SchrodingerB}, \cite{SchrodingerC}, unfortunately do not making the least counter-offer, that the  "exactly-defined (sharp) time" is the inconsequence inside of QM, and that moreover, so to speak, independently because of that, the special status of the time leads to an impediment in matching of QM with the relativity principle].

The natural question arises, why the remarks and indications of two from the three main founding fathers of the quantum theory were remained without any attention from the side of the researchers in the field.
It seems to be connected with Pauli’s conclusion, based on well known
his theorem, that the introduction of an time operator $\hat{T}$ must fundamentally be abandoned
and that the time $t$ in quantum mechanics has to be regarded to be an ordinary number
\cite{Pauli}. In other words, the consequence of Pauli’s theorem is the nonequality in rights of a time
coordinate in comparison with space coordinates for a description of quantum systems, which in its turn has led in the standard formulation of quantum theory to that, that the time is considered  to
be not a dynamical variable, but a mere parameter marking the evolution of a quantum
system, that is, the time is believed to be an external variable, which is independent on the
dynamics of any given system. 
Recently, the given theorem was reconsidered. In particular, it has been proved \cite{Galapon}, \cite{GalaponE},
that in quantum theory to classical variable "time", in contrast to the conclusion of Pauli, can be
put in the correspondence the self-adjoint time operator, like to space coordinates, energy,
impulse et cetera. Thus, the equality in rights of time coordinate and space coordinates was
reestablished. It is  also an additional confirmation, that the  Minkowski space is a single whole, that is, it is the 
 homogeneous physical object.

The semi-quantum-relativistic  character of Dirac canonical quantization method has expressed in its role on the introduction of the wave function of the photon and on the conception of the photon in electrodynamics itself (see below the brief description of the discussion on the notions of the photon wave function and the  photon conception to be the genuine particle and the arguments, allowing to crown the discussion in favour of the given notions).

 The canonical Dirac quantization method was developed in the work \cite{Dovlatova_Yerchuck} in
three aspects. The first aspect is its application the only to observable quantities. The second aspect
is the realization along with the well known semi-quantum-relativistic time-local quantization of the space-local quantization, which remains, however, also semi-quantum-relativistic being to be time-global. The main result is the development of the fully quantum-relativistic
space-time-local quantization method. It is the third aspect. In other words, it has been proved theoretically, that along with an indication of the time instant of photon creation (or annihilation) the space coordinate can be also determined.  Just the given result allows to finish the above indicated discussion about the possibility to describe the photons by wave function, which has been taking place over a long period of time, see below. It will be shown,inaccordance with aforesaid remark that the given problem, being to be emerging at the beginning of quantum physics era can be positively solved - photon state can be described by wave function, which can be built like to wave functions of another particles, for example like to the wave function of  the electron. 
The detailed argumentation of the given conclusion and the proof of the concept of the photon being  to be the genuine particle is the main aim of the given work.

\section{Concept of Photon Status}

Let us concern of the status of the photon in the modern physics at all. The researchers, dealing in the field of elementary particles, consider the photon to be the genuine particle to be self-evident, see, for instance, \cite{Djouadi}. Within the frames of the Standard Model of particle physics the photon is considered to be the messenger of the electromagnetic
interaction to which are subject charged particles. In other words, the interaction of electrically
charged particles is realised within the frames of the Standard Model through the exchange of photons \cite{Djouadi}. Let us remark, that the   Standard
Model, being to be the quantum and relativistic theory which describes in a unified framework the
electromagnetic, weak and strong forces of elementary particles is based in a very powerful principle, local or gauge symmetry: the
fields corresponding to the particles of a given internal symmetry group. In the given model  to each particle is associated a field that has a given number of degrees of
freedom. It is interesting, that according to the   Standard
Model the 
field associated to the massless photon has two degrees of
freedom \cite{Djouadi}. Let us cite the article \cite{Mishra} in the aspect , concerning the differences between  the messengers of the electromagnetic
interaction, that is, photons and the messengers of the weak force interaction, that is, W and Z bosons and their relation to Higgs boson: "The Higgs boson is a massive elementary particle predicted to exist by the Standard Model. It plays a
unique role in the Standard Model, by explaining why the other elementary particles are massive. In particular, the Higgs boson would explain why the
photon has no mass, while the W and Z bosons are very heavy. Elementary particle masses, and the differences between electromagnetism (mediated by the photon) and the weak force (mediated by the W and Z bosons), are critical to many aspects of the structure of microscopic (and hence macroscopic) matter. 
The CMS and ATLAS experiments at the Large
Hadron Collider (LHC) at CERN in Geneva reported the first experimental evidence of the Higgs boson's
existence on July 4, 2012. Subsequently, the Higgs boson mass has been measured to be about 125 GeV." It is seen from the given citate that the photon has the individual place among the genuine elementary particles in Standard Model and that it is considered to be self-evident, but its properties are quite different from the very heavy  $W$ and $Z$ bosons. Really,
in the concept of elementary particles existing at present
"the Higgs boson emerges and disappears by borrowing energy from and returning energy to
the particles in the electroweak interaction, respectively. Returning of energy from the
Higgs scalar boson to the particles is through the absorption of the Higgs scalar boson by the
particles. When a massless particle in the electroweak interaction absorbs the Higgs scalar boson,
the Higgs scalar boson becomes the longitudinal component of the massless particle, resulting
in the massive particle and the disappearance of the Higgs scalar boson" \cite{Chung}. According to \cite{Chung} "The observed Higgs boson at the LHC is a remnant of the Higgs boson. At the beginning
of the universe, all particles in the electroweak interaction were massless. The Higgs boson
appeared by borrowing energy symmetrically from all particles in the electroweak interaction.
The Higgs boson coupled with all massless particles, including leptons, quarks, and gauge bosons,
in the electroweak interaction. All massless particles except photon absorbed the Higgs boson
to become massive particles. The asymmetrical returning of energy from the Higgs boson by the
absorption of the Higgs boson is called the symmetrical breaking of the electroweak interaction
in the Standard Model. In the cases of  massive particles, including leptons, quarks, and
weak gauge bosons, the Higgs boson disappeared. In the case of massless photon the unabsorbed Higgs boson became the remnant of the Higgs boson. Being specific to the electroweak
interaction, the remnant of the Higgs boson could not return the borrowed energy to any other
massless particles". So it is established that the photons occupy a peculiar place regarding
the interaction with Higgs bosons. From here, taking into account the comparison with the W
and Z boson formation it has been done the   reazonable suggestion in that the
photons cannot be considered to be the gauge spin-1 bosons of electromagnetism.

So, the experts in the elementary particles' theory really consider the presence of the place of the photon  among the genuine elementary particles  to be self-evident, but its place is considered to be peculiar in the relation to the interaction with the Higgs boson.

At the same time the experts in the the quantum electrodynamics have the opposite opinion. Let us start from textbooks. Power and Kramers write quite directly, that the photon cannot be considered being to be the relativistic particle \cite{Power}, \cite{Kramers}. Correspondingly, concerning the wave functions of the photons, Power writes, that, strictly speaking, the wave functions of the photons are not existing. The main argument is the following. The fields $\vec{E}$ and $\vec{H}$ being to be satysfuing to Maxwell equations can be described according to his opinion by the real-defined functions. He concludes further, that they cannot be the solutions of the Schr\"{o}dinger equation, which is always the complex-defined function. We have to remark, that the given argument can be easily parried. It has been shown by the study of the symmetry of Maxwell equations, that the quantised EM-field is always described  the only by complex-defined field functions \cite{Dovlatova_Yerchuck}. In other words, they are the solutions of the  corresponding  Schr\"{o}dinger equation. Bohm \cite{Bohm}, in its turn, writes - strictly speaking, the function, describing the probability to find the light quantum, for instance, in space interval $(x, x+dx)$ does not exist. Really, the given conclusion is quite correct within the frames of semiclassical Dirac quantisation method over its space-global character, and it becomes to be incorrect by  the photon field description within the quantization method, proposed in \cite{Dovlatova_Yerchuck}, which has both the space and the time local character.
It is shown in  \cite{Dovlatova_Yerchuck}, that the operator functions  of creation $ \hat{a}^{+}_{\alpha}(z,t)$ and annihilation  $\hat{a}_{\alpha}(z,t)$ can be represented in the form
\begin{equation}
\label{eq69a}
\begin{split}
\hat{a}^{+}_{\alpha}(z, t) = \frac{1}{ \sqrt{ 2 \hbar \lambda_0  m_{\alpha} \omega_{\alpha}}} \left[ m_{\alpha} \omega_{\alpha}\hat {q}_{\alpha}(z, t) - i \hat {p}_{\alpha}(z, t)\right],
\end{split}
\end{equation}
\begin{equation}
\label{eq63a}
\begin{split}
&\hat{a}_{\alpha}(z, t) = \frac{1}{ \sqrt{ 2 \hbar \lambda_0  m_{\alpha} \omega_{\alpha}}} \left[ m_{\alpha} \omega_{\alpha}\hat {q}_{\alpha}(z, t) + i \hat {p}_{\alpha}(z, t)\right],
\end{split}
\end{equation}
by means of which the local space-time quantization of EM-field is realized.
The variables $\hat {q}_{\alpha}(z, t)$, $\hat {p}_{\alpha}(z, t)$ in \ref{eq69a} and \ref {eq63a} are canonically conjugated coordinate [amplitude of the normal mode, which have the dimensionality of the length] and impulse operator functions,corresponding to $\alpha$-mode of quantised EM-field, $\alpha\in N$,
The given functions satisfy the following relation. The value $\lambda_0$ is analogue of Planck constant $\hbar$. Although $\lambda_0$ and Planck constant $\hbar$ are equidimensional, however, their numerical
coincidence seems to be unobvious, since Planck constant characterizes the "seizure" of the
time by propagating of EM-field, while $\lambda_0$
characterises the "seizure" of the space. The value $\omega_{\alpha}$ is the circular frequency of EM-field $\alpha$-mode, $m_{\alpha}$ is the constant for the fixed mode, which have the dimensionality of the mass and which is introduced for the comparison with the  mechanical harmonic oscillator. 

The operator functions  $ \hat{a}^{+}_{\alpha}(z,t)$,   $\hat{a}_{\alpha}(z,t)$ in \ref{eq69a} and \ref {eq63a} satisfy to the relation
\begin{equation}
\label{eq72}
[\hat{a}_{\alpha}(z, t), \hat{a}^{+}_{\beta}(z, t)] = -i \delta_{\alpha\beta} \hat{e},
\end{equation}
where $\hat{e}$ is unit operator in the space of the  functions  $ \hat{a}^{+}_{\alpha}(z,t)$,   $\hat{a}_{\alpha}(z,t)$ representation.
Then the probability $dP(z,t)$ to find the light quantum, for instance, in space interval $(z, z+dz)$ and in time interval $(t, t+dt)$ is
\begin{equation}
\label{eq69av}
\begin{split}
dP(z,t)=
|\hat{a}^{+}_{\alpha}(z, t)|0\rangle|^2 dz dt,
\end{split}
\end{equation}
 where 
$|0\rangle$ is the vacuum state of EM-field.

Allow us to discuss thereupon the typical mistakes, which are widespread in the scientific literature on the quantum electrodynamics. In particular, in \cite{Scully} is insisted, that there are fundamental differences in the description of an photon to be the particle, despite the marvellous resemblance in motion equations for the photon and for the neutrino. Moreover the authors insist, that 
the description of an photon to be the particle is impossible. It contradicts very strongly  to the Einstein concept of light quanta to be the particles, introduced in the physics for the first time basing on experimental results on photoeffect, that is, with experimental grounds. Let us remark that the concept of light quanta to be the particles was considered already by Newton, although the only on the intuitive level without any experimental grounds. Let us consider "the arguments" in \cite{Scully}  in more details. The authors of \cite{Scully} argue in the following way. They write the relation for the plane wave, polarized in x-direction and propagating in z-direction in the form of
\begin{equation}
\label{eq1abz}
\phi(\vec{r},t) = 
 \vec{e}_x \frac {1}{\sqrt{V}}\exp[i(k_z z + \omega_{k_0} t)],
\end{equation}
where V is the volume of the propagation space, $k_z$ is the value of the  wave vector
$\vec{k_0} = k_z\vec{e}_z$, $\omega_k$ is the frequency, corresponding to $\vec{k_0}$.
In fact, the relation (\ref{eq1abz}) is incorrect, since instead of the scalar function in the left part of the relation has to be the vector function. However, it can be the only misprint. Further, the authors argue, that if to give the additional impulse in the direction $x$, that is, when a new wave vector $\vec{k}$ will be
\begin{equation}
\label{eq1abg}
\vec{k} = k_z\vec{e}_z + k_x\vec{e}_x,
\end{equation}
and to make the field transformation 
\begin{equation}
\label{eq1abm}
\exp (i k_x x)
\end{equation}
 a new   function they represent in the form
\begin{equation}
\label{eq1abh}
\tilde{\phi}(\vec{r},t) = 
 \vec{e}_x \frac {1}{\sqrt{V}}\exp[i(k_z z +  k_x x + \omega_k t)].
\end{equation}
In the given case, according to the opinion of the authors, the Maxwell equation
\begin{equation}
\label{eq1abf}
\nabla\cdotp\left[\begin{array} {*{20}c} \vec{\phi}(\vec{r},t) \\ \vec{\chi}(\vec{r},t) \end{array}\right] = 0
 \end{equation}
 is not true. They write
\begin{equation}
\label{eq1abk}
\nabla\cdotp\tilde{\phi} = \frac{\partial}{\partial x}[ \frac {1}{\sqrt{V}}\exp[i(k_z z +  k_x x + \omega_k t)]] \neq 0.
\end{equation}
From here the authors have concluded by the comparison with a unrelativistic particle with nonzero rest mass (with electron), that the representation of the photon to be the particle is erroneous.
All the  argumentation is incorrect. First, the transformation like to $\exp (i k_x x)$ [in general case $\exp (i \vec k \vec r)$] has to be applied to all field functions, that is, to the vector-function $\vec{\chi}(\vec{r},t)$ too, admittedly, $\vec{\chi}(\vec{r},t)$ is invariant in the particular case of the transformation $\exp (i k_x x)$ considered, since given  transformation is the rotation about y-axis, coinciding with the direction of the vector-function $\vec{\chi}(\vec{r},t)$ in the complex plane (ix, z), which can be set up to the real plane (x, z) by biective mapping. Second, the vector-function $\vec{\phi}(\vec{r},t)$ will be rotated on the angle $\theta$, determined by the relation $\tan\theta = \frac{k_x}{k_z}$, that is, it will be directed along a new unit vector $\vec{e}_{x'}$, at that,   $(\vec{e}_{x}\vec{e}_{x'}) = \cos\theta$.
Simultaneously, the transformation of $z$-axis also takes place $z \rightarrow z'$. The new direction of a z-axis, that is, z'-direction will coincide with  the new vektor $\vec{k}$ direction, at that, it is also a new impulse direction. The physical mistake of authors \cite{Scully} consist in that that they don't take into account that the propagation direction of the free EM-wave is determined by the impulse direction, that is, after the field functions transformation $\exp (i k_x x)$ EM-wave will be propagated along $z'$-direction. Consequently, instead of the relation (\ref{eq1abk}) has to be 

\begin{equation}
\label{eq1abwv}
\nabla\cdot \vec{\tilde{\phi}}(\vec{r},t) = \frac{\partial}{\partial x'}[ \frac {1}{\sqrt{V}}\exp[i(k_z' z'  + \omega_k t)]] = 0.
\end{equation}
in a full agreement with all Maxwell equations.

The authors of \cite{Scully} consider the second example, which, according to their opinion, is the most significant argument in the favour of the inapplicability of the representation of the photon to be the particle. The second example is two-quantum transitions. The amplitude of two-quantum detetection is \cite{Scully}

\begin{equation}
\label{eq1afb}
\Psi^{(2)}(\vec{r_1},t_1;\vec{r_2},t_2) = \langle 0|\hat{E}^{(+)}(\vec{r_2},t_2) \hat{E}^{(+)}(\vec{r_1},t_1),
\end{equation}
where $\hat{E}^{(+)}(\vec{r_1},t_1)$, $\hat{E}^{(+)}(\vec{r_2},t_2)$ are field annihilation operators, arguments of which are indicating on the probability of the annihilation of the photons in detectors, located in the points $\vec{r_1}$, $\vec{r_2}$ in the time moments $t_1$, $t_2$ correspondingly. It seems to be correct within the frames of an existing quantization procedure.[See,however, the work \cite{Dovlatova_Yerchuck}, the results of which allow to consider another way of looking into the task of two-quantum transitions]. It has been obtained in \cite{Scully} for the case, when the relaxation rates from the atomic level $\langle a|$ into the atomic level $\langle b|$, $\gamma_a$ and from the atomic level $\langle b|$ into the atomic level $\langle c|$, $\gamma_b$ are satisfuing to the relation $\gamma_a$ $\gg$ $\gamma_b$, the following expression
\begin{equation}
\label{eq1nbv}
\begin{split}
&\Psi^{(2)}(\vec{r_1},t_1;\vec{r_2},t_2) = \\
&\Psi_{\alpha}(\vec{r_1},t_1)\Psi_{\beta}(\vec{r_2},t_2)\Psi_{\beta}(\vec{r_1},t_1)\Psi_{\alpha}\vec{r_2},t_2)  \\
&\Psi_{\alpha}(\vec{r_i},t_i) = \frac{\mathcal{E}_{a}}{{\Delta r_i}} \Theta(t_i - \frac{\Delta r_i}{c})\times\\
&\exp [-\gamma_a(t_i - \frac{\Delta r_i}{c})]\exp [-i\omega_{ab}(t_i - \frac{\Delta r_i}{c})],
\\
&\Psi_{\beta}(\vec{r_i},t_i) = \frac{\mathcal{E}_b}{{\Delta r_i}} \Theta(t_i - \frac{\Delta r_i}{c})\times\\
&\exp [-\gamma_b(t_i - \frac{\Delta r_i}{c})]\exp [-i\omega_{bc}(t_i - \frac{\Delta r_i}{c})],
\end{split}
\end{equation}
where $\omega_{ab}$, $\omega_{bc}$ are the frequencies of the atomic transitions
$\langle a|$ $\rightarrow$ $\langle b|$ and $\langle b|$ $\rightarrow$ $\langle c|$ correspondingly, $\Delta r_i$ is the distance from an atom to i-th detector, i = 1,2, $\mathcal{E}_a$, $\mathcal{E}_b$ are constants, $\Theta(t_i - \frac{\Delta r_i}{c})$ is the Heaviside step-like function. From here the authors conclude on the emission of two independent photons, that allows to retain the concept of the photon to be the particle. 

At the same time, the authors of \cite{Scully} insist, that in the case  $\gamma_b$ $\gg$ $\gamma_a$ the concept of the photon to be the particle cannot be retained. They have obtained the following expression for the  amplitude of a two-quantum detetection in the given case
\begin{equation}
\label{eq1nbm}
\begin{split}
&\Psi^{(2)}(\vec{r_1},t_1;\vec{r_2},t_2) = \\
& \frac{-\kappa}{\Delta r_1 \Delta r_2}\exp [-i(\omega_{ac}+\gamma_a)(t_1 - \frac{\Delta r_1}{c})] \Theta(t_1 - \frac{\Delta r_1}{c})\times\\
&\exp [[-i(\omega_{bc}+\gamma_b)\{(t_2 - \frac{\Delta r_2}{c})-(t_1 - \frac{\Delta r_1}{c})\}]\times\\
&\Theta[(t_2 - \frac{\Delta r_2}{c})-(t_1 - \frac{\Delta r_1}{c})] + [1 \leftrightarrow 2],
\end{split}
\end{equation}
in which $\omega_{ac}$ is the  frequency of the atomic transition
$\langle a|$ $\rightarrow$ $\langle c|$, $\kappa$ is constant. It is the relation 1.5.44 in \cite{Scully}, from which the authors conclude, that both the events are strongly correlated and the representation of the photons to be the particles is incorrectly. 

At the same time, the mathematical structure of the relation (\ref{eq1nbm})  is equivalent to the the mathematical structure of the relation (\ref{eq1nbv}), if the constant $\kappa$ to represent in the form $\kappa$ = $\mathcal{E'}_a$ $\mathcal{E'}_b$. It can be factorized like to (\ref{eq1nbv}), however, the first transition instead of the frequency $\omega_{ab}$  has the frequency $\omega_{ac}$.
It physically means, that the process of  two-quantum transitions in a three level atom by $\gamma_b$ $\gg$ $\gamma_a$ is accompanying  by the emission of two photons with the energy corresponding to the distance between the first and the third levels and with the energy corresponding to the distance between the second and the third levels. It means, that the concept of the photons to be the particles can be retained in the given case too.

It is interesting to remark that the authors of \cite{Scully} themselves by setting forth the Weisskopf-Wigner theory of  the spontaneous emission of two-level atom write: "...the function
\begin{equation}
\label{eq1mbm}
\begin{split}
\Psi_{\gamma}(\vec{r},t) = \langle 0|\hat{E}^{(+)}(\vec{r},t)|\gamma_0\rangle,
\end{split}
\end{equation}
can be interpreted being to be a certain form of the photon wave function. It has been done in the analogue with the particle wave function". $|\gamma_0\rangle$ in  (\ref{eq1mbm}) is the EM-field state in the  point of time $t > 0$, which corresponds to the atom, localised at the $\vec{r}_0$ point, $ \langle 0|$ is the vacuum state of  EM-field, being to be  the state of  EM-field in the initial point of time $t = 0$, $\hat{E}^{(+)}(\vec{r},t)$ is the positive-frequency part of the operator of the electric component of EM-field.

Therefore, it is seen that the position of Scully and Zubairy is self-contradictory. They, insisting on the one hand on the inapplicability of the representation of the photon to be the particle have used on the second hand the notion of the photon wave function, at that its definition has been done  in the full analogue with the definition of the wave function of the well established particle with nonzeroth rest mass, for instance, the electron.

In the general case, the wave function of the photon can be obtained similarly  to the wave function of the usual particle, that is, it  is given by  the expression
 
\begin{equation}
\label{eq1mbq}
\begin{split}
\Psi(\vec{r},t) = \langle \vec{r}|\psi(t)\rangle,
\end{split}
\end{equation}
 representing itself the scalar product of the eigenvector of the position operator (cordinate part) and the vector of the state, giving the time-dependent part. The given representation is possible, since according to space-time local quantisation method, developed in \cite{Dovlatova_Yerchuck} along with an indication of the time instant of photon creation (or annihilation) the space coordinate for the given event can be also determined. The given theory is applicable, since the state $\langle\vec{r}|$ is represented through the creation operator $\hat{\psi}(\vec{r})$, which acting on the vacuum state $|0\rangle$ creates the particle in the space point $\vec{r}$ in accordance with the expression

\begin{equation}
\label{eq1mbt}
\begin{split}
\langle\vec{r}| = \langle \hat{\psi}(\vec{r})|0\rangle.
\end{split}
\end{equation}

So we have for the photon wave function the following expression

\begin{equation}
\label{eq1mbl}
\begin{split}
\Psi(\vec{r},t) = \langle 0|\hat{\psi}(\vec{r})| \psi(t)\rangle,
\end{split}
\end{equation}
coinciding in its form with the customary expression for the wave function of the particle in the matter.

 The theoretical consideration of the structure of a quantised EM-field has been done in \cite{DAA}. The fundamental result, obtained by Dirac, that  the dynamical system, which consists of the ensemble of identical bosons is equivalent to the dynamical system, which consists of the ensemble of oscillators, was used in \cite{DAA} to show, that the presence of the scalar  charge function $\rho(\vec{r},t)$, which was established in  \cite{Dovlatova_Yerchuck} and which is peer force scalar characteristic of an electromagnetic  field along with vector force characteristics $\vec{E}(\vec{r},t)$, $\vec{H}(\vec{r},t)$ agrees  with the charge neutrality of photons. The simplest analogue in its mathematical description in the physics of the condensed matter is the chain of bosonic (spin S = 1) carbon atoms in \textit{trans}-polyacetylene. It has been shown, that 
neutral photons are topological relativistic solitons with nonzero spin value, which is equal to $\frac{1}{2}$ instead of the prevalent viewpoint,
that the photons possess by spin $S = 1$.   
It was argued, by the way, that 
the representation of  photons to be the result of the spin-charge separation effect in the rest massless "boson-atomic" structure of EM-field  makes substantially more clear the nature of the corpuscular-wave dualism \cite{DAA}.

Let us reproduce the experimental confirmations of the model proposed, which are given in \cite{DAA}.
Very strong argument in favour of the model proposed is the well known experimentally
observed phenomenon of an electron-positron annihilation. It is well theoretically described, see,
for example \cite{Berestecky}, and experimentally confirmed that by the direct annihilation of electron-positron
pair two photons are produced. At the same time, the explanation for the case of the relative
velocity of annihilating particles near to zero, how can be produced from two particles with spin
value 1/2 also two particles, however with spin value 1, is in fact absent. It is evident, that
the boson model presented explains given disagreement between experimental data on electron-positron annihilation and the theoretical viewpoint on the photon to be the spin-1 particle in a
natural way - from two particles with spin values 1/2, that is from electron-positron pair can be
produced two photons the only with spin values 1/2.

The second argument is the existence of the dependence of a resonance microwave absorption
rate on the spin value of absorbing centers, which was found in electron spin resonance (ESR)
spectroscopy for the first time in \cite{Yerchak}, \cite{YerchakD}. It was established, that spin 1-centers absorb
microwave power with the rate exceeding the the absorption rate by spin 1/2-centers precisely
two times more. From given result it follows the reasonable suggestion, that if the photons were
possessing by spin 1 they couldn't be absorbed in ESR conditions by the centers with the spin
value 1/2 (like to the absence of any ESR-absorption by the centers with zeroth spin value).
The third argument is the aforedescribed peculiarities of interactions of the messengers of EM-interaction - photons'  with Higgs bosons within the frames of Standard Model in comparison with qualitatively quite different interactions of Higgs bosons with boson messengers $Z$ and $W$ of the weak interaction and the coincidence of the dimensionality of the 2D-spinor space, corresponding to the description of the particles with spin value, equaled to 1/2 (instead of 3D-vector space) with the dimensionality equaled to 2 aforeindicated, which is accepted for the description of the photon field in Standard Model. Ler us remark,In  that the spinor representation of quantized EM-field agrees well with the classical picture, in which the only two polarisation of EM-field  are experimentally observed.
It is given in the paper \cite{DAA}  the new physically clear interpretation of a corpuscular-wave dualism. It is explained by the  complex structure of EM-field. Really, since the quantized EM-field represents itself according to the model proposed the discrete massless boson-"atomic" structure like to an atomic structure in condensed matter the origin of wave
in a given structure is determined by the mechanism, quite analogous to the Bloch wave formation in the 
solid state of condensed matter. It is harmonic trigonometric functions for Maxwellian EM-field,
which determine its wave character. At the same time there
are simultaneously the corpuscules, propagating along given EM-field boson-"atomic" chain
structure, that is, chargeless spin 1/2 topological relativistic solitons - photons, formed in usual
conditions (or spinless charged solitons in so-called "doped" EM-field structure). It is concluded in \cite{DAA}, that the display of the corpuscular or wave nature of EM-field will
be dependent on experimental conditions. The experimental results on the interference and diffracion of light reported in \cite{Dempster} reported seem to be the excellent confirmation for the given conclusion. The observation of the only corpuscular properties at a low light intensity is easily explained by rectlinear propagation of photons, the size of which seems to be comparable with the size of the silver embryo. At the same time, the massless boson-atomic density  is rather low  in interslit space at  a low light intensity and the formation of Bloch-like waves do not take place. In fact, the given experiment indicates that there is threshold in the massless boson-atomic density for the formation of Bloch-like waves and correspondingly threshold in light intensities and the wave properties of the light can be observed the only by the intensities exceeding the given threshold. In other words owing to the ability of a quantum  Fermi liquid (like to any liquid) to spreading the infill of all interslit space takes place, however the concentration of masless bosons has to be sufficient to realize the interslit space infill. It is interesting to remark, that the representation of EM-field to be quantum liquid is well agree with classical representation of the propagation process of the light through small apertures, in which  the apertures are postulated being to be new light point sources. It is  quite similar to  spreading of any liquid through small apertures by the presence of  the pressure [the presence of light pressure is taken into account], for instance, like to spreading of a eau-de-Cologne from a bottle of eau-de-Cologne with a pulverizer in hairdressing saloons.

Therefore, the experimental results of  Dempster and Batho seem to be the most striking argument in the favour of the quantum  Fermi liquid model of EM-field, proposed in 
\cite{DAA}.

We have to remark that all existing in quantum optics theories do not  explain 
 correctly the phenomenon of the interference, in particular, the classical experiment of Yung with two slits. According to \cite{Scully}, the appearance of the interference in  Yung experiment depend on the coherence degree of two beams of light only  and do not depend on their intensity in contradiction with the results of \cite{Dempster}. All the more, the attempt to consider the propagation of a single photon through two slits simultaneously undertaken by some authors seems to be the grossest blunder.]

\section{Physics of Photon Absorption}

After recovering the status of photon in the electrodynamics to be the genuine particle and introducing some clarity into its geometry and spin-charge characteristics,  it seems to be significant to solve the task concerning the physical essence of the photon absorption process itself in the matter. The given task seems to be not trivial, and it was in fact formulated by Dirac already in 1927, however, it is not solved up to now, although the classical description of the absorption process of the energy of EM-field is well elaborated. Dirac in the work \cite{P.Dirac} writes, that the photon possesses by the strange peculiarity, it, seemingly, discontinues to exist, when it is located in one of its stationary states, just, in the zeroth state, in which its energy and its impulse are equal to zero. The absorption of a light quantum, according to  Dirac suggestion, is equivalent to a light quantum jump in the zeroth state, and its emission is its jump from the zeroth state in a some new state, where the photon existance is physically evident, so, it seems  that it was recreated. The given Dirac's comment can be really considered being to be the formulation of the task of quantum absorption process, at that Dirac remains  seemingly the alternative - the discontinuity of a photon existance in the zeroth state, or its revival by some way in the given state. However, the words "We have to suggest, that there are in the zeroth state infinitely many of the light quanta", - indicate that   Dirac  gives the  preference to the photon revival variant, which he uses in the subsequent computations of the quantum absorption and emission process, at that, he deals by computations with a great but a finite number of photons. The given conclusion indicates on a very deep insight of  Dirac
in the field. Futher we will represent the development of the given Dirac photon revival idea in the zeroth state and its experimental confirmation by modern spectroscopic studies.

There is also another aspect in the description of photon absorption and emission processes. It is the finite time of the interaction of photons with matter by their absorption and emission, which does not take into account both in classical and quantum electrodynamics. The attention was drawn on the given aspect also by Dirac \cite{P.Dirac}. Dirac has indicated, that for the correct description of the electrodynamic system the fact that forces propagate not instantaneous, but with the light velocity, has to be taken into consideration. Finite times of the interactions of EM-field with the matter have to be taken into account correctly by the studies of absorption, reflection and other processes of interactions of EM-field with the matter in all the stationary measurements. At the same time the Bloch equations are used both by quantum and classical descriptions of the given processes. The given equations take into consideration the relaxation of the only excited states by stationary processes. In other words the transition of absorbing centers (atoms , molecules and so on) into an excited state by the interaction with photons is postulated in implicit form to be instantaneous by all stationary optical and radiospectroscopy studies. At that, the level of stationary signals is 
considered to be determining by the interplay of the photon flux intensity and effectivity of the relaxation processes from the excited state.
At the same time quite another situation can be realized, in which the level of stationsry signals is determined by the time of the transition in the excited state, which can be rather long, characteristic times of which can even exceed the time of relaxation processes from the excited state. Let us give the physical ground for the given conclusion.
In the work \cite{Slepyan_Yerchak} was 
developed a theory of Rabi oscillations
in a periodical 1D chain of two-level quantum dots (QD) with tunneling
coupling, exposed to quantum light. The role of
interdot coupling and Rabi oscillations on each other was
considered in detail. The following conclusions have been done
from the studies.
1.The interdot tunneling in the QD chain
exposed to quantum light leads to the appearance of spatial
modulation of Rabi oscillations and in the appearance of the phenomenon of Rabi waves propagation.
It is shown, that Rabi waves can propagate if the
light mode wave vector has nonzero component along the
chain axis. Characteristics of the Rabi waves depend strongly
on relations between parameter of electron-photon coupling,
frequency deviation and transparency factors of potential
barriers for both of levels if individual QDs.
2.Traveling Rabi wave represent the quantum state of
QD chain dressed by radiation, that is, joined states of electron-hole (e-h)
pair and photons. The qualitative distinction of these states
from the similar states of single dressed atom is the space-time
modulation of dressing parameter according to the traveling
wave law. The propagation of traveling Rabi wave
looks like supported by periodically inhomogeneous nonreciprocal
effective media, whose refractive index is deter
mined by electric field distribution. For quantum description of the give Rabioscillation propagation process the authors introduce the  quasiparticles of a new type - rabitons, which can be consideredbeing to be the 
generalization of Hopfield polaritons51,53 for the case of indirect
quantum transitions.
3.Two traveling Rabi modes with different frequencies
of Rabi oscillations corresponds to the given value of wave number.
The range of Rabi oscillation frequencies is limited by the
critical value, which different for both of the modes. The QD chain
is opaque in the regime of Rabi oscillation frequencies below
the critical value. At the same time it is shown, that the critical frequencies and dispersion
characteristics of Rabi modes depend on number of photons.
4.The formation of different types of Rabi wave packets was considered. It has been found, that they represent themselves
 arbitrary superpositions of four partial subpackets with
different amplitudes, frequency shifts, and velocities of a motion.
Two of subpackets correspond to the contribution of
excited initial state and two others caused by the ground
initial state contribution. It was established that Rabi wave packets transfer energy,
inversion, quasimomentum, electron-electron, and electron-photon
quantum correlations along the chain. The number of
subpackets can be diminished in specific circumstances.
5.For the case of QD chain considered in the work cited,it is found, that Rabi oscillations qualitatively change the electron tunneling
picture in the given chain. In contrast to the case of the absence of
electron-photon coupling, the movement of initially groundstate
subpacket is governed by tunneling transparency of excited
energy level and vice versa. Thus, Rabi oscillations can
stimulate the tunneling through low-energy level and suppress
it through high-energy one.
6.It has been established, that Rabi wave packet movement along the QD chain alters
the light statistics. Particularly, it was predicted for the QD chain, exposed
to coherent light the drastic modification
of the standard collapse-revival phenomenon: collapses and
revivals appear in different areas
of the chain space.It seems especially significant the conclusion of the authors of \cite{Slepyan_Yerchak}, that the phenomenon of
Rabi waves'formation can take place in a number of other distributed
systems strongly coupled with electromagnetic field.
For the example they indicate on the possibility of
Rabi waves'formation in  superconducting circuits based on Josephson
junctions, which are currently the most experimentally advanced
solid-state qubits. According to the opinion of the authors of \cite{Slepyan_Yerchak} the
qubit-qubit capacitance coupling in the chain of qubits
placed inside a high-Q transmission-line resonator will be
responsible for the Rabi waves propagation similar to described in the paper above cited.

We wish to draw attention, that the analysis of the results obtained in the work \cite{Slepyan_Yerchak} allow to predict the additional new quantum phenomenon -  the two stage process of a photon absorption, which can have rather long times, even more long than the relaxation times of the excited atomic states. So, in the first stage the rabiton formation takes place. The given process can be rather fast. The second stage is determined by lifetime of rabitons, in which photons, being to be not absorbed coexist with absorbing matter excitons. It seems to be  substantial for the practical applications, that the given state is quantum coherent state. It also significant, that photons in the given bound state are moving withrather small velocities in comparison with $c$, that is, with the light velocity in vacuum.  The extrapolation of the rabiton velocity to zero means that photon can exist in the state with zero energy and zero impulse in full correspondence with Dirac guess. In fact, pinning of photons by absorbing centers is taking place. Which charactristics remains for the zero mass photons in the given pinned state. It is spin, which in correspondence with results of \cite{Dovlatova_Yerchuck} is the most fundamental characteristic of quantum states.
We wish also draw attention that the creation and annihilation operators used in quantum field theory and in quantum electrodynamics in particular operate with creation (annihilation) of the energy and impulse and they are not concerned spin of photons. In fact, the content of the given operators (at least in an application to photons) can by slighly modified. They  transfer photons from pinned state and in the pinned state correspondingly.  The agreement of quantum field calculations, which do not touch the spin of photons with experiments seems to be the strong argument for the given conclusion.

We have to remark that the existance of pinned photons in condensed matter does not contradict the special relativity theory, since according to it the photons have the velocity c in any inertial system being to be propagated in vacuum.
Understanding of the physics of absorption processes with rabiton formation allows to predict the spectroscopiic experiments in which the signal level for a  stationary state will be dependend on the the times of the transfer of a system of absorbing centers into excited state, moreover to measure the given times. If the   
the times of the transfer of a system of absorbing centers into an excited state are more long in comparison with  at least one from relaxation times which are responsible to the transfer from an excited state,  than the signal will have unsaturating behavior in dependence on a radiation power, whereas its maximum will be registered in phase being be shifted on a some angle in the correspondence with reference signal by the  modulation method of an absorption detection, if the modulation frequency is near to the reverce value of  the time of the transfer of a system of absorbing centers into excited state. The  modulation method of an absorption detection is used in the  magnetic resonance spectroscopy, although it is applied sometimes in the optical spectroscopy too.

The corresponding time can be measured by the method, analogous to the method, which is very good developed in photoconductivity studies \cite{Ryvkin}.
Let us gine the brief descriptionof the given method in the application to  magnrtic resonance experiments. The static magnetic field  will be can be replaced in  the linear approximation by the modulation of the microwave field. Then its intensity I(t)   can be described by the relation
\begin{equation}
\label{eq6ab}
\begin{split}
I(t) = I_a (1-\cos \omega_m t),
\end{split}
\end{equation},
where $\omega_m$ is the value of the modulation frequency.
The phenomenological equation for the linear responce A(t) of the system of absorbinng centers on the given modulation is

\begin{equation}
\label{eq7ab}
\begin{split}
\frac{dA(t)}{dt} = \alpha I_a (1-\cos \omega_m t) - \frac{A(t)}{\tau_g},
\end{split}
\end{equation}
where $\tau_g$ is characteristic time of growth of the responce signal, $\alpha$ is the value characterizing electron-photon effective interaction, suggested to be constant. The solution of the given equation is

\begin{equation}
\label{eq8ab}\begin{split}
&A(t) = \alpha I_a \tau_g + \alpha I_a\tau_g\frac{1}{1+\omega^2_m\tau^2_g}\times\\
&(\tau_g\omega\sin\omega_m t 
+ \cos \omega_m t) + A_0exp(- \frac{t}{\tau_g})\end{split},
\end{equation}
where, $A_0$ is the constant of integration. Taking into acount the initial conditions A(t) = 0 at t=0, we will have
\begin{equation}
\label{eq9ab}\begin{split}
&A(t) = \alpha I_a \tau_g [ 1 - \frac{2 +\omega^2_m\tau^2_g}{1 +\omega^2_m\tau^2_g}]exp(- \frac{t}{\tau_g}) + \\
&\frac{\alpha I_a\tau_g}{1 +\omega^2_m\tau^2_g}(\tau_g\omega\sin\omega_m t + \cos \omega_m t)
\end{split}.
\end{equation}

Then in the stationary conditions we will have

\begin{equation}
\label{eq10ab}
\begin{split}
&A_{st}(t) =   \frac{\alpha I_a\tau_g}{1 +\omega^2_m\tau^2_g}(\tau_g\omega\sin\omega_m t \\
&+ \cos [\omega_m t -\arctan(\omega_m \tau_g)]
\end{split}.
\end{equation}
Therefore, for the angle $\vartheta$ between the  maximum of the registered absorption signal and reference signal by sinchronous detection we obtain the expression
\begin{equation}
\label{eq11ab}
\vartheta =  \arctan(\omega_m \tau_g);
\end{equation}
which allows to determine $\tau_g$.

We have analysed the literature and have found the corresponding results. It has been reported in the work \cite{Ertchak}, that the paramagnetic absorption changes linearly independence on microwave power in the boron implanted
(425-150 keV) diamond films (Figure 9 in \cite{Ertchak}). At the same time the EPR signal recorded in these films and in the nickel-implanted diamond single crystals (335 MeV, $5\times{10^{14}} cm^{-2}$) \cite{Ertchak}, \cite{Ertchak_Stelmakh} has an additional angle in respect to the phase of the hf modulation field
(Figure15 in \cite{Ertchak}). In the diamond films this angle is anisotropic and changes within 20 degrees when samples are rotated in a
static magnetic field. At the same time in the nickel-implanted diamond single crystals the additional phase angle was found to be isotropic.
It is characteristic that the kinetics of the intensity of an EPR signal recorded in phase $I(H_1)$ and
quadrature $ I_q(H_1)$  with the hf modulation field is the same, that is,  linear in the boron-implanted 425-150 keV films
(Figure 9 in \cite{Ertchak}) and superlinear in the nickel-implanted  diamond SCs (Figure 7 in \cite{Ertchak_Stelmakh}). On the one
hand, this fact is evidence that the EPR lines recorded both in phase and in quadrature with the hf modulation
field correspond to the same PCs, thereby confirming again the appearance of an additional phase angle in an EPR
signal recorded with the use of hf modulation. On the other hand, this points to the fact that in the case considered the
mechanism of the formation of paramagnetic absorption, which is responsible for the EPR signal recorded in quadrature with
the hf modulation field, differs from the known mechanism. Usually, the EPR signal recorded in quadrature with
the hf modulation field corresponds to PCs having long times of paramaguetic relaxation and differing in nature
from PCs, being to be  responsible for the EPR signal recorded in phase \cite{Weger}. So it was found the direct proof, that the amplitude and phase of stationary EPR absorption in the samples above indicated are determined by the times of the interaction of absorbing centers with microwave range photons, indicating in its turn on the quantum character of EPR absorption in the samples studied in \cite{Ertchak}. The value of the time of transfer in excited state for the angle $\vartheta$ = 20 degrees is
$\tau_g$ = $5.8\times{10^{-7}} s$. It is direct indication that in the samples studied in \cite{Ertchak}, \cite{Ertchak_Stelmakh} the longlived coherent states are realized. 

The similar very interesting results were reported  in the paper \cite{Shaltiel}. The authors write: "A complication in
the measuring process has been observed. It is related to variations in the signal phase that
may develop through the measuring process. The reference AC phase of the lock-in-getector (LID) is adjusted
for usual EPR measurements to yield the optimum paramagnetic resonance signal. However
in case of the microwave dissipation observed in anisotropic superconductors this
adjustment turns out to be not appropriate. Variations in the signal phase were observed
when varying the magnitude of the variables involved  (DC magnetic field,
temperature, sample orientation and others). This effect necessitates a procedure where the
lock-in detector phase has to be adjusted, throughout the measuring process, to be in-phase
with the AC phase. To overcome this problem, conducted during the measuring process, the
measurements were performed in steps of the variables involved, and at each step the lock-in
detector phase had to be adjusted accordingly. This procedure results in an extremely
cumbersome measurement, where the possibilities offered by the modern EPR Bruker
spectrometer ELEXSYS E500 are extremely useful. A feature of this spectrometer is its
possibility to measure the signal intensity in steps of two variables. It allows obtaining the
measured signal intensity at each step, where the lock-in phase is varied from zero to 180 degrees.
Repeating the measurement, by changing, in steps, the magnitude of the variable involved
(magnetic field, temperature, angle of the oriented crystal) gives the measured
signal intensity being to be the function of the variable at the full 180 degrees range of the lock-in phase. Then
the measurements were analyzed using a proper computer program". It allowed  to authors  to obtain
two curves, the measured signal amplitude, and the corresponding signal phase, being to be the 
function of the corresponding variable involved ( DC magnetic field, AC magnetic field, sample
orientation angle and temperature). Thus, the actual “calculated” signal was derived from
the measured signal amplitude and its corresponding signal phase.

$Bi_2Sr_2CaCu_2O_{8+\delta}$ (Bi2212) single crystal was studied, zero field cooled
 to 4K with the DC magnetic field applied parallel to the a-b plane.  The measured
signal amplitude exhibits two maxima followed by an exponential decay towards zero at
high fields. The signal phase is close to zero at low fields and drops steeply towards -180 degrees at
high fields. It indicates that the signal is in phase and out of phase with the AC field at low
and high fields, respectively. The deviations of the phase shift, that is deviations (20 degrees) from 0 degrees at
low fields and from 180 degrees at high fields have been found, however they have not been investigated so far and need according to the opinion of authors in further
attention. The authors give the only comment that the value of phase shift from low to high fields which is  less than 180°
cannot be explained just by a constant shift of the calibration.  The comparison of the given very interesting result with the results described in \cite{Ertchak}, \cite{Ertchak_Stelmakh} leads  to the conclusion that the nature of the appearance of the angle between the maximum of the signal registered and reference signal can be  entirely the same. In other words, the level of stationary signals in Bi2212 single crystal seems to be governed by the relaxation time, connected with transfer of its electronic system into excited state, that is, with finite time of the photons' absorption. To confirm the given interpretation the study of the  signal dependence on the  microwave power level is necessary (the given dependence was not represented in the work cited).

It is clear  that the dynamics of spectroscopic transitions cannot be described in the systems above described by known Bloch equations or Torrey equations both in classical and semiclassical forms even qualitatively, since the process of the photon absorption (or emission) is suggested in Bloch equations or Torrey equations in implicit form being to be instantaneous. It raises the concernment of the works \cite{Yearchuck_Yerchak_Dovlatova}, \cite{QFTDST}, where the equations for  the dynamics of spectroscopic transitions are represented, which are appropriate for the case ofthe processes of the photon absorption (or emission) with any characteristic times.

The system of difference-differential equations presented in  the the works \cite{Yearchuck_Yerchak_Dovlatova}, \cite{QFTDST} can describe 
the dynamics of spectroscopic transitions for both radio
and optical spectroscopy for the model, representing itself
the 1D-chain of N two-level equivalent elements coupled
by exchange interaction (or its optical analogue for the
optical transitions) between themselves and interacting with
quantized EM-field and quantized phonon field. Naturally, the equations are true for any 3D system of
paramagnetic centers  or optical centers by the absence
of exchange interaction.
In comparison with semiclassical
description, where the description of dynamics of spectroscopic
transitions is exhausted by one vector equation (Landau–Lifshitz equation or Landau–Lifshitz
based equation), by quantum–electrodynamic consideration the  Landau–Lifshitz type
equation describes the only one subsystem of three-part system — photon
field- electronic subsystem of the matter - phonon field. The complete system of difference-differential equations include the matrix equations for  the photon
field and for the  phonon field. By the way, let us remark, that it has been shown in \cite{Yearchuck_Yerchak_Dovlatova}, that Landau–Lifshitz equation is fundamental physical
equation underlying along with the dynamics of a  magnetic moment motion, the dynamics of spectroscopic transitions and
transitional phenomena, however, let us accentuate once again, that its use is sufficient the only by classical or semiclassical description of the interaction of EM-field with matter.

\section{Conclusions}
The status of the photon in the modern physics was analysed. It is ascertained, that within the frames of the Standard Model of particle physics the photon is considered to be the genuine elementary particle, being to be the messenger of the electromagnetic
interaction to which are subject charged particles. In contrast, the analysis of the scientific literature and textbooks on quantum electodynamics has shown, that the experts in quantum electodynamics (in particular, in quantum optics) insist, that 
the description of an photon to be the particle is impossible anf that the photon wave function does not exists. The given viewpoint was carefully analysed and its falseness was proved. The expression for a photon wave function is presented. So, the status of the photon in
quantum electodynamics was restored. 

The physics of quantum absorption process is analysed. It is argued in accordance with Dirac guess, that the photon revival take place by its absorption. Being to be a soliton, it seems to be keeping safe after an energy absorption in a pinned state, possessing the only by spin. 

It is shown, that the time  of the transfer of absorbing systems in an excited state is finite in general case and moreover, that it can govern the stationary signal  registered. The given result seems to be  significant for all the  stationary spectroscopy, in which at present the transfer of absorbing systems in an excited state is considered to be instantaneous. 
The value of the time of transfer in excited state in the samples studied in \cite{Ertchak}, \cite{Ertchak_Stelmakh} was evaluated tobe equal
$\tau_g$ = $5.8\times{10^{-7}} s$. It is accentuated, that it is direct indication that in the samples studied in \cite{Ertchak}, \cite{Ertchak_Stelmakh} the longlived coherent states are realized, which in its turn indicates on the practical significance of the given systems for the  modern technology.

\end{document}